\documentstyle[12pt]{article}

\newcounter{eq}
\newcounter{sc}

\def\overleftrightarrow#1{\vbox{\ialign{##\crcr
 $\leftrightarrow$\crcr\noalign{\kern-1pt\nointerlineskip}
 $\hfil\displaystyle{#1}\hfil$\crcr}}}

\newlength{\minitwocolumn}
\begin{document}

\begin{flushright}
DPUR/TH/72\\
October, 2021\\
\end{flushright}
\vspace{20pt}

\pagestyle{empty}
\baselineskip15pt

\begin{center}
{\large\bf  Scale Invariance and Dilaton Mass
\vskip 1mm }

\vspace{20mm}

Ichiro Oda\footnote{
           E-mail address:\ ioda@sci.u-ryukyu.ac.jp
                  }

\vspace{10mm}
           Department of Physics, Faculty of Science, University of the 
           Ryukyus,\\
           Nishihara, Okinawa 903-0213, Japan\\

\end{center}


\vspace{10mm}
\begin{abstract}

We consider a generic scale invariant scalar quantum field theory and its symmetry breakdown.
Based on the dimension counting identity, we give a concise proof that dilaton is exactly massless 
at the classical level if scale invariance is broken spontaneously.
On the other hand, on the basis of the generalized dimension counting identity, we prove that 
the dilaton becomes massive at the quantum level if scale invariance is explicitly broken 
by quantum anomaly.

It is pointed out that a subtlety occurs when scale invariance is spontaneously broken through a scale invariant 
regularization method where the renormalization scale is replaced with the dilaton field. In this case, the dilaton 
remains massless even at the quantum level after spontaneous symmetry breakdown of scale symmetry,
but when the massless dilaton couples non-minimally to the Einstein-Hilbert term and is applied for cosmology, 
it is phenomenologically ruled out by solar system tests unless its coupling to matters is much suppressed 
compared to the gravitational interaction. 

\end{abstract}

\newpage
\pagestyle{plain}
\pagenumbering{arabic}


\section{Introduction}

A dilaton is defined as a Nambu-Goldstone boson of scale (or dilatation) invariance and appears ubiquitously 
in various areas of modern theoretical physics such as string theory \cite{Polchinski} and scalar-tensor 
gravity \cite{Fujii}. Since scale symmetry prohibits the appearance of dimensionful parameters in an action, 
it is not a symmetry in our world. Thus, in any case scale invariance must be broken spontaneously 
or explicitly by quantum anomaly.  

One of the important problems associated with the dilaton is the magnitude of its mass. If the dilaton is exactly 
massless, it mediates a long-range force between massive objects as in the Newtonian force. This fact imposes 
a phenomenological constraint on parameters in a theory including the dilaton such as the Brans-Dicke model \cite{Brans}
since the long-range force stemming from the massless dilaton could affect the perihelion advance of Mercury, 
for instance.  On the other hand, if the dilaton has a small mass, there might be an observational possibility 
of a non-Newtonian force whose force range depends on the magnitude of the mass and the detail 
of a model \cite{Fujii2}.

The aims of this article are two-fold. One aim is related to a recent paper discussing 
a no-go theorem that a massless dilaton cannot exist in renormalizable, unitary scalar quantum field theory 
(QFT) in four dimensions \cite{Nogradi}. In this article, we wish to present an alternative and very concise proof
of this no-go theorem on the basis of the dimension counting identity and its anomalous generalization.   
One of attractive points in our proof is that we can deal with non-renormalizable QFT as well.

The other aim is to point out a subtlety of the massless dilaton when it is applied for cosmology. In the
manifestly scale invariant regularization method \cite{Englert}-\cite{Ghilencea}, scale invariance is spontaneously 
broken while maintaining quantum scale invariance, that is, no quantum anomaly. In this case, the dilaton is precisely 
massless even at the quantum level, but in order to carry out the scale invariant regularization method, there appear
some problems to be solved in future, those are, non-renormalizability \cite{Shaposhnikov2}, fine-tuning of coupling 
constants \cite{Kugo} and treatment of the unbroken phase \cite{Tamarit}. In this article, we wish to point out 
that there is another problem when the massless dilaton is applied for cosmology since the scalar force generated by 
the massless dilaton could in general affect the solar system tests and the late-time cosmology, so it must be screened 
by the still-unknown mechanism in order to be consistent with observations and predictions obtained from general relativity.
 
We close this section with an overview of this article. In Section 2, we discuss classical scale invariance
and show that the dilaton mass as well as the cosmological constant are identically vanishing at the
stationary point of a potential.
In Section 3, we work with a quantum theory with scale anomaly. It is then easy to show that the dilaton,
which is massless at the classical level, becomes massive owing to the anomaly.
In Section 4, we consider the case where scale invariance is spontaneously broken and quantum scale invariance
is maintained via the scale invariant regularization method. 
The final section is devoted to the conclusion.

\section{Classical scale invariance} 

In this section, we will present a concise proof that the dilaton is precisely massless when 
a theory has strict scale invariance at the classical level. This proof is based on the dimension counting 
identity of a potential and holds for both classical and quantum scale invariances. However, as will be argued 
in a later section, one has to pay additional attention to the case of the quantum scale invariance. 

Let us start with a classically scale invariant theory with $n$ real scalar fields 
$\phi_i ( i = 1, 2, \cdots, n )$\footnote{We use a Minkowskian metric of signature $(- + + +)$.}:
\begin{eqnarray}
{\cal{L}} = - \frac{1}{2} \sum_{i=1}^n \partial_\mu \phi_i \partial^\mu \phi_i -  V ( \phi ),
\label{CSI-model}
\end{eqnarray}
where $V ( \phi )$ is a generic scale invariant potential of scalar fields $\phi_i$. 
The classical scale invariance then leads to the dimension counting identity:
\begin{eqnarray}
\sum_{i=1}^n \phi_i \frac{\partial V ( \phi )}{\partial \phi_i} =  4  V ( \phi ).
\label{DC-iden}
\end{eqnarray}

At the stationary point of the potential $\phi_i = v_i$, we have
\begin{eqnarray}
\left. \frac{\partial V ( \phi )}{\partial \phi_i} \right|_{\phi = v}  =  0.
\label{Stat-point}
\end{eqnarray}
Together with Eq. (\ref{Stat-point}), Eq. (\ref{DC-iden}) implies that the vacuum energy, 
or equivalently the cosmological constant when the theory couples to gravity, is vanishing 
at the classical level: 
\begin{eqnarray}
V ( v ) =  0.
\label{Zero-CC}
\end{eqnarray}

Taking the partial derivative of Eq. (\ref{DC-iden}) with respect to the scalar fields $\phi_j$,
we obtain 
\begin{eqnarray}
\sum_{i=1}^n \phi_i \frac{\partial^2 V ( \phi )}{\partial \phi_i \partial \phi_j} 
=  3 \frac{\partial V ( \phi )}{\partial \phi_j}.
\label{DC-iden2}
\end{eqnarray}
At the stationary point of the potential, this equation reads
\begin{eqnarray}
\sum_{j=1}^n \left. \frac{\partial^2 V ( \phi )}{\partial \phi_i \partial \phi_j} \right|_{\phi = v}  v_j =  0.
\label{Mass-eq}
\end{eqnarray}
In a similar manner, one can obtain a series of equations holding at the stationary point
\begin{eqnarray}
\sum_{j, k}  \left. \frac{\partial^3 V ( \phi )}{\partial \phi_i \partial \phi_j \partial \phi_k} \right|_{\phi = v}  
v_j v_k &=&  0,
\nonumber\\
\sum_{j, k, l}  \left. \frac{\partial^4 V ( \phi )}{\partial \phi_i \partial \phi_j \partial \phi_k \partial \phi_l} \right|_{\phi = v}  
v_j v_k v_l &=&  0,
\nonumber\\
\sum_{m}  \left. \frac{\partial^5 V ( \phi )}{\partial \phi_i \partial \phi_j \partial \phi_k \partial \phi_l \partial \phi_m} \right|_{\phi = v}  
v_m &=&  0, 
\label{Series-eq}
\end{eqnarray}
and so on.

At this point, let us focus our attention on Eq. (\ref{Mass-eq}). The existence of a nonzero solution $v_i \neq 0$
leads us to the equation
\begin{eqnarray}
\det \left. \frac{\partial^2 V ( \phi )}{\partial \phi_i \partial \phi_j} \right|_{\phi = v} =  0.
\label{Det}
\end{eqnarray}
For simplicity, let us take $v_i$ to
\begin{eqnarray}
v_i =  ( 0, 0, \cdots, 0, v).
\label{VEV}
\end{eqnarray}
Then, Eq. (\ref{Mass-eq}) states that
\begin{eqnarray}
\left. \frac{\partial^2 V ( \phi )}{\partial \phi_i \partial \phi_n} \right|_{\phi = v} =  0.
\label{V-matrix}
\end{eqnarray}
Since we assume that the potential $V ( \phi )$ is a generic one in the sense that 
\begin{eqnarray}
\left. \frac{\partial^2 V ( \phi )}{\partial \phi_I \partial \phi_J} \right|_{\phi = v} \neq  0,
\label{generic-V}
\end{eqnarray}
with being $I, J = 1, \cdots, n-1$, we can obtain the result
\begin{eqnarray}
{\rm rank} \left. \frac{\partial^2 V ( \phi )}{\partial \phi_i \partial \phi_j} \right|_{\phi = v} =  n - 1.
\label{Rank}
\end{eqnarray}

We can now make clear the meaning of Eq. (\ref{Rank}) by expanding the potential around $v_i$ as follows:
\begin{eqnarray}
V (\phi) = V (v) + \sum_i \left. \frac{\partial V ( \phi )}{\partial \phi_i} \right|_{\phi = v}  \varphi_i
+ \frac{1}{2} \sum_{i, j} \left. \frac{\partial^2 V ( \phi )}{\partial \phi_i \partial \phi_j} \right|_{\phi = v}  
\varphi_i \varphi_j + \cdots,
\label{Talor}
\end{eqnarray}
where $\dots$ describes higher-order terms in $\varphi_i$ and we have assumed that
\begin{eqnarray}
\varphi_i \equiv \phi_i - v_i, \qquad | \varphi_i | \ll 1.
\label{Varphi}
\end{eqnarray}
With the help of Eqs. (\ref{Stat-point}) and (\ref{Zero-CC}), Eq. (\ref{Talor}) can be cast to the form
\begin{eqnarray}
V (\phi) &=& \frac{1}{2} \sum_{i, j = 1}^n \left. \frac{\partial^2 V ( \phi )}{\partial \phi_i \partial \phi_j} \right|_{\phi = v}  
\varphi_i \varphi_j + \cdots
\nonumber\\
&=& \frac{1}{2} \sum_{I = 1}^{n - 1} \left. \frac{\partial^2 V ( \phi )}{\partial \phi_I^2} \right|_{\phi = v}  
\varphi_I^2 + \cdots,
\label{Talor2}
\end{eqnarray}
where in the last equality the symmetric matrix $\left. \frac{\partial^2 V ( \phi )}{\partial \phi_i \partial \phi_j} 
\right|_{\phi = v}$ with the rank (\ref{Rank}) is transformed to an $(n - 1) \times (n - 1)$ diagonal matrix by a
suitable orthogonal matrix.\footnote{The sufficient condition such that the stationary point $\phi_i = v_i$ is the local
minimum of a potential is that the Hessian matrix $\left. \frac{\partial^2 V ( \phi )}{\partial \phi_I \partial \phi_J} 
\right|_{\phi = v} (I,  J = 1, 2, \cdots, n-1)$, is the positive definite matrix.}   
The final expression in Eq. (\ref{Talor2}) clearly shows that spontaneous symmetry breakdown (SSB) 
of scale invariance occurs, thereby yielding $n - 1$ massive scalar fields and one massless scalar field, 
which is nothing but the dilaton predicted by the Nambu-Goldstone (NG) theorem. It is of interest that all massless 
scalar fields except for the dilaton become massive as a result of the SSB of scale invariance.

Now we will present two simple examples, one of which has a property (\ref{generic-V}) and we therefore 
obtain $n - 1$ massive scalar fields and one massless dilaton after the SSB of scale invariance 
whereas the other example has a potential such that there is an internal $SO(n)$ global symmetry, 
which is spontaneously broken to an $SO(n-1)$ symmetry, and consequently 
we have $n - 1$ and $1$ Nambu-Goldstone bosons corresponding to the SSB of the internal symmetry 
and scale invariance, respectively.

The first example has a potential:   
\begin{eqnarray}
V ( \phi ) =  \frac{1}{2} \sum_{I, J = 1}^{n-1} \lambda_{IJ} \phi_I \phi_J \phi_n^2,
\label{S-pot}
\end{eqnarray}
where the coupling constant is symmetric, $\lambda_{IJ} = \lambda_{JI}$.
Then, the first derivatives of the potential become 
\begin{eqnarray}
\frac{\partial V ( \phi )}{\partial \phi_I} =  \sum_{J = 1}^{n-1} \lambda_{IJ} \phi_J \phi_n^2,
\qquad
\frac{\partial V ( \phi )}{\partial \phi_n} =  \sum_{I, J = 1}^{n-1} \lambda_{IJ} \phi_I \phi_J \phi_n.
\label{S-pot-der}
\end{eqnarray}
Note that we can see that Eq. (\ref{DC-iden}) is certainly satisfied in this case.

As a nontrivial vacuum which triggers the SSB of scale invariance, let us take a stationary point
\begin{eqnarray}
( \phi_I, \phi_n ) =  ( 0, 0, \cdots, 0, v).
\label{S-Vac}
\end{eqnarray}
At the vacuum (\ref{S-Vac}), the second derivatives of the potential read
\begin{eqnarray}
\left. \frac{\partial^2 V ( \phi )}{\partial \phi_I \partial \phi_J} \right|_{\phi = v} 
&=& \left. \lambda_{IJ} \phi_n^2 \right|_{\phi = v} =  \lambda_{IJ} v^2,  \qquad
\left. \frac{\partial^2 V ( \phi )}{\partial \phi_I \partial \phi_n} \right|_{\phi = v} 
= \left. 2 \sum_{J = 1}^{n-1} \lambda_{IJ} \phi_J \phi_n \right|_{\phi = v} = 0,
\nonumber\\
\left. \frac{\partial^2 V ( \phi )}{\partial \phi_n^2} \right|_{\phi = v} 
&=& \left. \sum_{I, J = 1}^{n-1} \lambda_{IJ} \phi_I \phi_J \right|_{\phi = v} = 0. 
\label{S-Vac-Value}
\end{eqnarray}
Using this result, we find that Eqs. (\ref{generic-V}) and (\ref{Rank}) are valid 
so we can obtain $n - 1$ massive scalar fields and one massless dilaton after the SSB of scale invariance.

Of course, this result can be checked directly by expanding the fields about the vacuum (\ref{S-Vac}) as
\begin{eqnarray}
\phi_I = \varphi_I, \qquad  \phi_n = v + \varphi_n, 
\label{S-Expan}
\end{eqnarray}
by which the potential (\ref{S-pot}) can be rewritten as
\begin{eqnarray}
V ( \phi ) &=&  \frac{1}{2} v^2 \sum_{I, J = 1}^{n-1} \lambda_{IJ} \varphi_I \varphi_J 
+ v \sum_{I, J = 1}^{n-1} \lambda_{IJ} \varphi_I \varphi_J \varphi_n 
+ \frac{1}{2} \sum_{I, J = 1}^{n-1} \lambda_{IJ} \varphi_I \varphi_J \varphi_n^2.
\label{S-pot-exp}
\end{eqnarray}
This expression means that we have $n - 1$ massive scalar fields $\varphi_I$ 
with the mass matrix squared $M_{IJ}^2 = v^2 \lambda_{IJ}$ and a massless dilaton 
$\varphi_n$.  

Next let us consider the second example whose potential is of form:
\begin{eqnarray}
V ( \phi ) =  \frac{\lambda}{4 !} \left( \sum_{i=1}^n \phi_i^2 - \phi_{n+1}^2 \right)^2,
\label{O-pot}
\end{eqnarray}
where $n + 1$ massless scalar fields are introduced, one of which has negative sign in 
front of the $\phi_{n+1}^2$ term whereas the other $n$ fields have positive sign and give us 
an $SO(n)$ internal symmetry.

The stationary condition is defined as in Eq. (\ref{Stat-point}) and the cosmological constant is vanishing owing to
scale invariance as in Eq. (\ref{Zero-CC}). Using Eq. (\ref{O-pot}), we can also check explicitly the validity
of the identity (\ref{DC-iden}). Furthermore, Eq. (\ref{DC-iden2}) yields
\begin{eqnarray}
\sum_{i=1}^{n+1} \phi_i \frac{\partial^2 V ( \phi )}{\partial \phi_i \partial \phi_j} 
=  3 \frac{\partial V ( \phi )}{\partial \phi_j}.
\label{DC-iden2-O}
\end{eqnarray}
At the stationary point of the potential, this equation takes the form
\begin{eqnarray}
\sum_{j=1}^{n+1} \left. \frac{\partial^2 V ( \phi )}{\partial \phi_i \partial \phi_j} \right|_{\phi = v}  v_j =  0.
\label{Mass-eq-O}
\end{eqnarray}

As before, the existence of a nonzero solution $v_i \neq 0$ provides us with Eq. (\ref{Det}). 
As a stationary point, let us take $v_i$ to
\begin{eqnarray}
v_i \equiv ( v_1, v_2, \cdots, v_n, v_{n+1}) =  ( 0, 0, \cdots, 0, v, v).
\label{VEV-O}
\end{eqnarray}
Then, on account of Eqs. (\ref{O-pot}) and (\ref{VEV-O}), it turns out that the rank of the matrix 
in Eq. (\ref{Mass-eq-O}) is given by
\begin{eqnarray}
{\rm rank} \left. \frac{\partial^2 V ( \phi )}{\partial \phi_i \partial \phi_j} \right|_{\phi = v} =  1.
\label{Rank-O}
\end{eqnarray}
This result tells us that we have only one massive particle and the remaining $n$ particles become massless
after SSB. In other words, the vacuum (\ref{VEV-O}) triggers the SSB of an $SO(n)$ symmetry 
to an $SO(n-1)$ symmetry in addition to the scale symmetry, and consequently we have $n$ massless NB bosons,
one of which is the dilaton generated by the SSB of scale invariance and the other $n - 1$ NB bosons 
come from that of the internal symmetry.

The above result can be also investigated by expanding the fields around the nontrivial vacuum (\ref{VEV-O})
obeying the stationary condition as follows. Taking the variation of the potential (\ref{O-pot}) with respect to 
$\phi_i$ with $i = 1, 2, \cdots, n$ and $\phi_{n+1}$, respectively, yields
\begin{eqnarray}
\frac{\partial V ( \phi )}{\partial \phi_i} &=& \frac{\lambda}{3 !} \phi_i \left( \sum_{i=1}^n \phi_i^2 - \phi_{n+1}^2 \right),
\nonumber\\
\frac{\partial V ( \phi )}{\partial \phi_{n+1}} &=& - \frac{\lambda}{3 !} \phi_{n+1} \left( \sum_{i=1}^n \phi_i^2 - \phi_{n+1}^2 \right),
\label{O-pot-var}
\end{eqnarray}
from which we can see the existence of a flat direction defined as
\begin{eqnarray}
\sum_{i=1}^n \phi_i^2 - \phi_{n+1}^2 = 0.
\label{Flat-direc}
\end{eqnarray}

Since the nontrivial vacuum configuration (\ref{VEV-O}) is a solution to Eq. (\ref{Flat-direc}), we can expand the fields
around this configuration as follows:
\begin{eqnarray}
\phi_I = \varphi_I, \qquad  \phi_n = v + \varphi_n, \qquad  \phi_{n+1} = v + \varphi_{n+1},
\label{Expan}
\end{eqnarray}
where $I = 1, 2, \cdots, n-1$. Then, the potential (\ref{O-pot}) can be cast to the form
\begin{eqnarray}
V ( \phi ) &=&  \frac{\lambda v^2}{6} ( \varphi_n - \varphi_{n+1} )^2 
+ \frac{\lambda v}{6} ( \varphi_n - \varphi_{n+1} ) \left( \sum_{i=1}^n \varphi_i^2 - \varphi_{n+1}^2 \right)
\nonumber\\
&+& \frac{\lambda}{4 !} \left( \sum_{i=1}^n \varphi_i^2 - \varphi_{n+1}^2 \right)^2.
\label{O-pot-exp}
\end{eqnarray}
Introducing new fields defined as
\begin{eqnarray}
\chi_n =  \frac{\varphi_n + \varphi_{n+1}}{\sqrt{2}}, \qquad
\chi_{n+1} =  \frac{\varphi_n - \varphi_{n+1}}{\sqrt{2}},
\label{New scalars}
\end{eqnarray}
the potential (\ref{O-pot-exp}) can be rewritten as
\begin{eqnarray}
V ( \phi ) =  \frac{\lambda v^2}{3} \chi_{n+1}^2 
+ \frac{\sqrt{2} \lambda v}{6} \chi_{n+1} \left( \sum_{I=1}^{n-1} \varphi_I^2 + 2 \chi_n \chi_{n+1} \right)
+ \frac{\lambda}{4 !} \left( \sum_{I=1}^{n-1} \varphi_I^2 + 2 \chi_n \chi_{n+1} \right)^2.
\label{O-pot-exp2}
\end{eqnarray}
Note that at the same time the kinetic term can be expressed in terms of the new fields 
in a canonical form:
\begin{eqnarray}
- \frac{1}{2} \sum_{i=1}^{n+1} ( \partial_\mu \phi_i )^2
= - \frac{1}{2} \sum_{I=1}^{n-1} ( \partial_\mu \varphi_I )^2
- \frac{1}{2} ( \partial_\mu \chi_n )^2 - \frac{1}{2} ( \partial_\mu \chi_{n+1} )^2.
\label{Kinetic term}
\end{eqnarray}
As a result, we have one massive scalar field $\chi_{n+1}$ with mass squared $M_{\chi_{n+1}}^2
= \frac{2}{3} \lambda v^2$, a massless dilaton $\chi_n$ and $n - 1$ massless NG bosons $\varphi_I$
corresponding to the SSB from an $SO(n)$ to an $SO(n-1)$, as expected. 

To close this section, it is worthwhile to point out the following facts obtained above: In classically
scale invariant theories, before the SSB of scale invariance, all particles are massless. Once scale
invariance is spontaneously broken, the massless particles become massive except for the dilaton
which is a NB boson corresponding to the SSB of scale invariance. Provided that there are  
internal symmetries in addition to scale invariance in a classical action, we have additional NB
bosons associated with the SSB of the internal symmetries. In this case, all particles become
massive but the NB bosons and the magnitude of mass is proportional to the scale $v$ of the
SSB.

\section{Quantum anomaly}

In this section, we work with quantum field theory (QFT) where we encounter a quantum anomaly called scale anomaly. 
It is well known that the presence of the ultra-violet divergences in QFT makes it necessary to introduce a dimensionful
renormalization scale $\mu$ in a theory, which could violate classical scale invariance. Provided that the renormalization
scale $\mu$ is introduced, the dimension counting identity must be generalized to
\begin{eqnarray}
\left( \mu \frac{\partial}{\partial \mu} + \sum_{i=1}^n \phi_i \frac{\partial}{\partial \phi_i} \right)
V ( \phi ) =  4  V ( \phi ),
\label{Anom-DC-iden}
\end{eqnarray}
where the dimension counting identity picks up an anomalous piece corresponding to a derivative
with respect to the renormalization scale $\mu$.

Now, at the stationary point $\phi_i = v_i$ obeying Eq. (\ref{Stat-point})\footnote{Precisely speaking,
in QFT the stationary point should be understood as the vacuum expectation value of the field operator,
i.e., $\langle \phi_i \rangle = v_i$, and a classical potential must be replaced by an effective potential.}, 
the cosmological constant is not zero and given by
\begin{eqnarray}
V (v) = \frac{\mu}{4} \left. \frac{\partial V ( \phi )}{\partial \mu} \right|_{\phi = v} \neq 0.  
\label{Q-CC}
\end{eqnarray}

Taking the partial derivative of Eq. (\ref{Anom-DC-iden}) with respect to the scalar fields $\phi_j$
produces\footnote{Here, for simplicity, we assume that the renormalized field $\phi_i$ is independent 
of $\mu$, i.e., the anomalous dimension is zero, which holds at one loop level.}
\begin{eqnarray}
\sum_{i=1}^n \phi_i \frac{\partial^2 V ( \phi )}{\partial \phi_i \partial \phi_j} 
=  3 \frac{\partial V ( \phi )}{\partial \phi_j} - \mu \frac{\partial^2 V ( \phi )}{\partial \mu \partial \phi_j}.
\label{Q-DC-iden2}
\end{eqnarray}
At the stationary point, this equation becomes
\begin{eqnarray}
\sum_{j=1}^n \left. \frac{\partial^2 V ( \phi )}{\partial \phi_i \partial \phi_j} \right|_{\phi = v}  v_j 
=  - \left. \mu \frac{\partial^2 V ( \phi )}{\partial \mu \partial \phi_i} \right|_{\phi = v}.
\label{Q-Mass-eq}
\end{eqnarray}
Since the RHS is not in general vanishing because of scale anomaly, for nonvanishing $v_i$ we must have
the equation:
\begin{eqnarray}
\det \left. \frac{\partial^2 V ( \phi )}{\partial \phi_i \partial \phi_j} \right|_{\phi = v} \neq  0.
\label{Q-Det}
\end{eqnarray}

Then, since the matrix $\left. \frac{\partial^2 V ( \phi )}{\partial \phi_i \partial \phi_j} \right|_{\phi = v}$ 
is a symmetric matrix and can be therefore diagonalized by an orthogonal matrix, Eq. (\ref{Q-Det})
implies that
\begin{eqnarray}
{\rm rank} \left. \frac{\partial^2 V ( \phi )}{\partial \phi_i \partial \phi_j} \right|_{\phi = v} =  n.
\label{Rank-Q}
\end{eqnarray}
Eq. (\ref{Rank-Q}) shows that all the massless scalar fields become massive after symmetry breaking of scale invariance
owing to the scale anomaly. To put it differently, we have no more a massless dilaton under the presence of the 
quantum anomaly at the quantum level although the theory was scale invariant at the classical level.

In order to verify this fact explicitly, let us consider an example with a single massless scalar field whose
Lagrangian density is of form:
\begin{eqnarray}
{\cal{L}} &=& - \frac{1}{2} \partial_\mu \phi \partial^\mu \phi -  V ( \phi )  
\nonumber\\
&=& - \frac{1}{2} \partial_\mu \phi \partial^\mu \phi - \frac{\lambda}{4 !} \phi^4,
\label{phi4-model}
\end{eqnarray}
which is manifestly invariant under scale transformation.

By following Coleman and Weinberg \cite{Coleman} and using the modified minimal-subtraction (or $\overline{MS}$ scheme), 
we calculate the effective potential to one loop level:
\begin{eqnarray}
V_{eff} (\phi) = \frac{\lambda}{4 !} \phi^4 + \frac{\lambda^2}{256 \pi^2} \phi^4 
\left( \ln \frac{\lambda \phi^2}{2 \mu^2} - \frac{3}{2} \right),
\label{CW}
\end{eqnarray}
where $\mu$ is the renormalization scale. It is easy to see that this effective potential satisfies the 
generalized dimension counting identity (\ref{Anom-DC-iden}) in the sense that
\begin{eqnarray}
\left( \mu \frac{\partial}{\partial \mu} + \phi \frac{\partial}{\partial \phi} \right)
V_{eff} ( \phi ) =  4  V_{eff} ( \phi ).
\label{CW-DC-iden}
\end{eqnarray}

Taking the derivatives of the effective potential over $\phi$ leads to
\begin{eqnarray}
\frac{\partial V_{eff} ( \phi )}{\partial \phi} &=& \lambda \phi^3 \left[ \frac{1}{3 !} 
+ \frac{\lambda}{64 \pi^2} \left( \ln \frac{\lambda \phi^2}{2 \mu^2} - 1 \right) \right],
\nonumber\\
\frac{\partial^2 V_{eff} ( \phi )}{\partial \phi^2} &=& \frac{\lambda}{2} \phi^2 \left[ 1
+ \frac{\lambda}{32 \pi^2} \left( 3 \ln \frac{\lambda \phi^2}{2 \mu^2} - 1 \right) \right].
\label{CW-Deriv}
\end{eqnarray}
From these equations, we can find $\langle \phi \rangle = v$ which minimizes the effective potential,
and the dilaton mass squared:
\begin{eqnarray}
v &=& \mu \sqrt{\frac{2}{\lambda}} \exp\left( { \frac{3 \lambda - 32 \pi^2}{6 \lambda} } \right),
\nonumber\\
m_D^2 &=& \left. \frac{\partial^2 V_{eff} ( \phi )}{\partial \phi^2} \right|_{\phi = v} 
= \frac{\lambda^2 v^2 }{32 \pi^2}.
\label{v-mass}
\end{eqnarray}
Note that since $v \propto \mu$, the dilaton mass is proportional to the renormalization scale 
$\mu$, which describes the magnitude of a violation of scale invariance.

\section{Quantum scale invariance}

It is a familiar fact that some fields have built-in mechanisms to keep them massless; vector fields 
have gauge invariance and fermions have chiral invariance. Perhaps, it is scale invariance that 
would make scalar fields be massless, but the existence of the ultra-violet divergences in QFT forces us 
to introduce a dimensionful renormalization scale $\mu$ in a theory, which violates classical scale invariance
inevitably at the quantum level. Without scale invariance, scalar fields would have no immunity to protect them 
from acquiring a nonzero mass.  

However, if there were no scale anomaly at the quantum level, we would also have the dimension counting 
identity (\ref{DC-iden}). Then, by following our argument in Section 2, we find that the dilaton remains massless 
even at the quantum regime as well as the classical one. In fact, Englert et al. \cite{Englert} have 
already proposed an interesting approach such that 
scale invariance can be maintained even at the quantum level if we introduce a scalar field, which was called 
``dilaton'', instead of the renormalization scale $\mu$. Their reasoning is so simple that let us follow it briefly: 
By using the dilaton, the Einstein-Hilbert action in general relativity, which has no scale anomaly, can be transformed 
to a locally scale invariant scalar-tensor gravity \cite{Dirac, Deser} \footnote{A relation between the locally scale 
invariant scalar-tensor gravity and Weyl transverse gravity was examined in Ref. \cite{Oda-C} and the Higgs mechanism 
in scale invariant gravitational theories has been investigated in Ref. \cite{Oda-H}.}, 
which could have a Weyl anomaly corresponding to local scale 
invariance.\footnote{Here the difference between global and local scale invariances is not important since any 
globally scale invariant theories can be promoted to local ones by coupling gravity in an appropriate manner.}  
With the order reversed, if we start with the locally scale invariant scalar-tensor gravity and take the unitary gauge, 
i.e., a constant dilaton, we reach the Einstein-Hilbert action. This quantum equivalence between general relativity 
and the locally scale invariant scalar-tensor gravity then suggests that the anomaly associated with local scale 
invariance, which exists in the scale invariant scalar-tensor gravity, might be fake since there is no such an anomaly 
in general relativity.  

It is true that the approach advocated by Englert et al. \cite{Englert} is indeed free from the quantum anomaly 
and we can explicitly check that the dilaton is exactly massless even at the quantum level \cite{Ghilencea}.
However, in this approach, which we will call the scale invariant regularization method, there are several problems 
we have to pay the price of success. First, such scale invariant theories are non-renormalizable as in general relativity 
and make sense only as effective field theories \cite{Shaposhnikov2}. Second, there is a fine-tuning problem for keeping
a flat direction stable against radiative corrections, which can be understood as the incarnation of the cosmological 
constant problem \cite{Kugo}. Finally, it is not clear how to treat with the unbroken phase ($\langle \phi \rangle = 0$ where $\phi$ 
is a dilaton) since the formalism makes sense only in the broken phase ($\langle \phi \rangle \neq 0$) \cite{Tamarit}.
In this section, we wish to point out that there is another phenomenological problem when we attempt to apply 
the scale invariant regularization method for cosmology in addition to the above three problems. 

It is natural that the massless dilaton $\phi$ couples to the Einstein-Hilbert term with the non-minimal coupling term\footnote{We 
follow the conventions and notation of the MTW textbook \cite{MTW}.}
\begin{eqnarray}
\frac{1}{\sqrt{-g}} {\cal{L}} = \frac{1}{2} \xi \phi^2 R - \frac{1}{2} \epsilon g^{\mu\nu} 
\partial_\mu \phi \partial_\nu \phi + L_m,
\label{ST-Lagr}
\end{eqnarray}
where $\epsilon = \pm 1 = \frac{|\omega|}{\omega}$ with $\omega$ being the Brans-Dicke parameter which
will appear shortly. Here $\epsilon = 1$ corresponds to a normal field with a positive energy while
$\epsilon = - 1$ does to a ghost field with a negative energy. Moreover, $\xi$ is a constant and 
in particular the case of $\xi = \frac{1}{6}$ and $\epsilon = - 1$ produces a locally scale invariant scalar-tensor gravity,
and $L_m$ denotes the Lagrangian for a generic matter field without mass terms. Finally, it is worthwhile to stress 
that the Langrangian density (\ref{ST-Lagr}) is invariant under scale transformation.

By defining a new scalar field $\varphi$, which is sometimes called the Brans-Dicke scalar field, 
and the Brans-Dicke parameter $\omega$ as  
\begin{eqnarray}
\varphi = \frac{1}{2} \xi \phi^2,  \qquad \omega = \frac{\epsilon}{4 \xi},
\label{phi-omega}
\end{eqnarray}
we can rewrite (\ref{ST-Lagr}) as a well-known Brans-Dicke Lagrangian:   
\begin{eqnarray}
\frac{1}{\sqrt{-g}} {\cal{L}} = \varphi R - \omega \frac{1}{\varphi} g^{\mu\nu} 
\partial_\mu \varphi \partial_\nu \varphi + L_m.
\label{BD-L}
\end{eqnarray}
The field equation for the Brans-Dicke scalar field then turns out to be
\begin{eqnarray}
\Box \varphi = \frac{1}{4 \omega + 6} T,
\label{BD-scalar}
\end{eqnarray}
where $T$ denotes the trace part of the energy-momentum tensor $T_{\mu\nu}$ for matter
field defined as $T_{\mu\nu} \equiv - \frac{2}{\sqrt{-g}} \frac{\delta (\sqrt{-g} L_m)}{\delta g^{\mu\nu}}$. 

In the manifestly scale invariant regularization method, scale invariance is expected to
be broken by Planck-scale physics \cite{Ghilencea}, so the dilaton will acquire a large vacuum expectation value (VEV),
$\langle \phi \rangle \sim M_{Pl}$ where $M_{Pl}$ is the reduced Planck mass defined in terms of the
Newton constant $G$ as $M_{Pl} \equiv \frac{1}{\sqrt{8 \pi G}}$. Indeed, this expectation can be realized by
the following argument: First, let us expand the dilaton $\phi$ about its VEV as
\begin{eqnarray}
\phi = \langle \phi \rangle + \tilde \phi,
\label{Exp-BD-dilaton}
\end{eqnarray}
and substitute this expansion into the Lagrangian (\ref{ST-Lagr}), from which we can obtain
\begin{eqnarray}
\frac{1}{\sqrt{-g}} {\cal{L}} = \frac{1}{2} \xi \langle \phi \rangle^2 R + \xi \left( \langle \phi \rangle \tilde \phi 
+ \frac{1}{2} \tilde \phi^2 \right) R
- \frac{1}{2} \epsilon g^{\mu\nu} \partial_\mu \tilde \phi \partial_\nu \tilde \phi + L_m,
\label{ST-Lagr2}
\end{eqnarray}
Next, to make the first term on the RHS of Eq. (\ref{ST-Lagr2}) coincide with the Einstein-Hilbert term,
we have to choose $\xi \langle \phi \rangle^2 = M_{Pl}^2$. With $\xi = {\cal{O}}(1)$, we have 
$\langle \phi \rangle \sim M_{Pl}$. This completes our confirmation that scale invariance is broken
around the Planck scale.\footnote{Incidentally, as will be seen in Eq. (\ref{Cassini}), $\xi$ is a tiny quantity.
If we take $\xi = 10^{-6}$ for instance, we have $\langle \phi \rangle = 10^3 M_{Pl}$, which is 
beyond the Planck mass so it is not physically interesting.} 

Eq. (\ref{BD-scalar}) implies that the Brans-Dicke scalar field mediates a long-range force between 
massive objects as in the Newtonian force and its coupling strength is given by $\frac{1}{4 \omega + 6}$,
which is roughly of the same order of magnitude as that of the Newtonian force since $\omega = \frac{\epsilon}{4 \xi}
= {\cal{O}}(1)$, that is, $\omega$ is of the order of unity. Note that as in the Brans-Dicke theory, most of models 
related to the dilaton have couplings with matters which are of gravitational strength.  

On the other hand, the currently best time delay measurements by the Cassini probe in the solar system 
gives us the bound \cite{Bertotti}
\begin{eqnarray}
\omega > 4.3 \times 10^3 \ (\mbox{or} \quad \xi < 5.8 \times 10^{-5}), \qquad \epsilon = 1.
\label{Cassini}
\end{eqnarray}
Unfortunately, the magnitude of the Brans-Dicke parameter $\omega$ is too large to be no room that 
it is close to anywhere around the unity. This problem is known as a long-standing curse in the Brans-Dicke theory
and its generalizations \cite{Fujii}. To solve this problem, some people have advocated that the dilaton might have small mass 
in such a way that it mediates a finite-range force \cite{Fujii2}. If its Compton wave-length is smaller than 
the size of the solar system, we can neglect the finite-range force from the massive dilaton, thereby making 
the constraint (\ref{Cassini}) be irrelevant phenomenologically.

However, with quantum scale invariance, the dilaton is exactly massless at the both classical and quantum levels, 
so it would be not impossible but very difficult to construct a realistic cosmological model which is consistent with 
the Cassini bound (\ref{Cassini}).

\section{Conclusions}

In this article, on the basis of the dimension counting identity and its generalization, we have given a concise proof 
that the dilaton is exactly massless at the classical level if scale invariance is broken spontaneously whereas it becomes 
massive at the quantum level if scale invariance is explicitly broken by quantum anomaly.
Moreover, we have shown that when quantum scale invariance is maintained through the scale invariant regularization
method, the dilaton remains massless at the quantum level. Compared to the proof in Ref. \cite{Nogradi}, our proof is simpler 
and has an advantage of being able to apply for non-renormalizable quantum field theory.

It is of interest to notice that before the spontaneous symmetry breakdown of scale invariance, all the particles are
massless because of scale invariance, and after the SSB they become massive except for the dilaton unless there are 
internal symmetries such as $SO(n)$.  Here all masses arise from a nonzero vacuum expectation value of the dilaton field. 

Furthermore, we have pointed out that in addition to three problems of the manifestly scale invariant
regularization scheme where the renormalization scale is replaced with a dilaton field, there is
the fourth problem which emerges when we couple the massless dilaton to the Einstein-Hilbert term 
in the non-minimal way; a difficulty of constructing a realistic cosmological model by using the massless dilaton.  
It seems that a crucial point is to develop a new mechanism for nullifying the scalar force stemming from the dilaton.  
We wish to find such a mechansim in future.\footnote{While preparing for this paper, an approach for screening
dilaton interaction was proposed in \cite{Burgess}.}

\begin{flushleft}
{\bf Acknowledgements}
\end{flushleft}

This work is partly supported by the JSPS Kakenhi Grant No. 21K03539.


\end{document}